\documentclass{article}\usepackage[hidelinks]{hyperref}

\usepackage{arxiv}

\usepackage[utf8]{inputenc} % allow utf-8 input
\usepackage[T1]{fontenc}    % use 8-bit T1 fonts
\usepackage{hyperref}       % hyperlinks
\usepackage{url}            % simple URL typesetting
\usepackage{booktabs}       % professional-quality tables
\usepackage{amsfonts}       % blackboard math symbols
\usepackage{nicefrac}       % compact symbols for 1/2, etc.
\usepackage{microtype}      % microtypography
\usepackage{lipsum}		% Can be removed after putting your text content
\usepackage{graphicx}
\usepackage[square,numbers]{natbib}
\usepackage{doi}

\usepackage{graphicx}
\usepackage[utf8]{inputenc}
\usepackage[english]{babel}
\usepackage{cancel}
\usepackage{amsmath}
\usepackage{url}
\usepackage{multirow}
\usepackage{array,multirow}
\usepackage{algorithm}
\usepackage{algorithmic}
\usepackage{graphicx}
\usepackage{verbatim}
\graphicspath{ {./figure/} }
\usepackage{babel}
\usepackage{xfrac}
\usepackage{dblfloatfix}
\usepackage{etoolbox}
\makeatletter
\patchcmd{\@makecaption}
  {\scshape}
  {}
  {}
  {}
\makeatother
\usepackage{xcolor}
\usepackage{tabularx}
\usepackage{amssymb}
\usepackage{amsthm}
\usepackage{array, boldline, makecell, booktabs}

\newtheorem{definition}{Definition}

\newtheorem{theorem}{Theorem}
\usepackage{multirow}
\usepackage{enumitem}
\usepackage{diagbox}
\usepackage{afterpage}

\usepackage{color,soul}
\usepackage{array,multirow}
\usepackage{caption}

\usepackage{microtype}
\usepackage{booktabs}
\usepackage[format=hang,font={small,bf}]{caption}
\usepackage[utf8]{inputenc}
\usepackage{color,soul}
\setul{0.5ex}{0.3ex}
\definecolor{Green}{rgb}{0,1,0}
\setulcolor{Green}
\usepackage[bottom]{footmisc}

\usepackage{nccmath}
\usepackage{mathtools}

\title{Discovering Structural Hole Spanners in Dynamic Networks via Graph Neural Networks}
\author{Diksha Goel\textsuperscript{1}, 
Hong Shen\textsuperscript{2}, 
Hui Tian\textsuperscript{3}, 
Mingyu Guo\textsuperscript{1}
\And
\normalfont 1. School of Computer Science, University of Adelaide, Australia\\\vspace{0.1in}
\{diksha.goel, mingyu.guo\}@adelaide.edu.au\\
2. School of Computer Science and Engineering, Sun Yat-sen University, Guangzhou, China\\\vspace{0.1in}
shenh3@mail.sysu.edu.cn\\
3. School of Information and Communication Technology, Griffith University, Australia\\
hui.tian@griffith.edu.au}

\date{}

\begin{document}
\maketitle
\begin{abstract}
Structural Hole (SH) theory states that the node which acts as a connecting link among otherwise disconnected communities gets positional advantages in the network. These nodes are called Structural Hole Spanners (SHS). SHSs have many applications, including viral marketing, information dissemination, community detection, etc. Numerous solutions are proposed to discover SHSs; however, most of the solutions are only applicable to static networks. Since real-world networks are dynamic networks; consequently, in this study, we aim to discover SHSs in dynamic networks. Discovering SHSs is an NP-hard problem, due to which, instead of discovering exact $k$ SHSs, we adopt a greedy approach to discover top-$\textbf{\textit{k}}$ SHSs. Motivated from the success of Graph Neural Networks (GNNs) on various graph mining problems, we design a Graph Neural Network-based model, GNN-SHS, to discover SHSs in dynamic networks, aiming to reduce the computational cost while achieving high accuracy. We analyze the efficiency of the proposed model through exhaustive experiments, and our results show that the proposed GNN-SHS model is at least 31.8 times faster and, on an average 671.6 times faster than the comparative method, providing a considerable efficiency advantage.
\end{abstract}

\keywords{Structural hole spanners; graph neural network; dynamic networks; pairwise connectivity.}

\section{INTRODUCTION}
The emergence of large-scale networks has inspired researchers to design new techniques to analyze and study the properties of these large-scale networks. The structure of the network inherently possesses a community structure where the nodes within the community share close interests, characteristics, behaviour, and opinions \cite{zannettou2018origins}. 
The absence of connection between different communities in the network is known as \textbf{\textit{Structural Hole (SH)}} \cite{burt2009structural}. The structural hole theory states that the users who fill the “\textit{holes}” between various users or groups of users that are otherwise disconnected get positional advantages in the network, and these users are known as \textbf{\textit{Structural Hole Spanners}} \cite{lou2013mining}. Figure 1 depicts Structural Hole Spanner in the network. 

Various solutions \cite{lou2013mining, he2016joint, xu2019identifying, tang2012inferring, rezvani2015identifying, xu2017efficient, ding2016method, zhang2020finding, luo2020detecting, goel2021maintenance, goel2024effective} are developed to discover SHSs in static networks. Nevertheless, the real-world network changes over time. For example, on Facebook and Twitter, links appear and disappear whenever a user friend/unfriend others on Facebook or follow/unfollow others on Twitter. As a consequence of the dynamic nature of network, discovered SHSs also change, and it is important to quickly discover new SHSs in the updated network. Classical SHS identification algorithms are considerably time-consuming and may not work efficiently for dynamic networks. In addition, the network may have already been changed by the time classical algorithms recompute SHSs. Hence, we need a fast mechanism that can efficiently discover SHSs as the network evolves continuously. 

\textbf{\textit{Graph Neural Networks (GNNs)}} \cite{thekumparampil2018attention, kipf2016semi} are deep learning-based techniques designed for graph problems. GNNs have shown remarkable results on graph optimization problems, due to which it has become a widely adopted graph investigation technique. In this paper, we study GNN for discovering SHSs in dynamic networks. %Therefore, we examine GNNs for finding SHS in dynamic networks.

 \begin{figure}[t!]
 \centering
  \includegraphics[width=0.37\columnwidth]{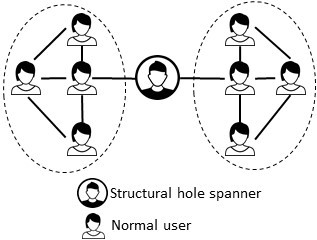}
  \caption{Structural Hole Spanner in the network.}
  \label{fig:SHS}
 \end{figure}

In literature, Goel et al. \cite{goel2021maintenance} proposed a solution for discovering SHSs in dynamic networks. The authors have considered one decremental update at a time. In contrast, real networks are highly dynamic where decremental as well as incremental updates take place. The network structure changes continuously; new nodes or edges either join or leave the network. In our model, we consider both incremental as well as decremental edge updates of the network and propose  a model \textbf{\textit{GNN-SHS}} (\textit{\underline{G}raph \underline{N}eural \underline{N}etworks for discovering \underline{S}tructural \underline{H}ole \underline{S}panners in dynamic networks}), a GNN-based framework to discover SHSs in the dynamic network. We regard the dynamic network as a sequence of snapshots and aim to discover SHSs in these snapshots. Our proposed  GNN-SHS model uses the network structure and features of nodes to learn embedding vectors of nodes. Our model aggregates embedding vectors from the neighbors of the nodes, and the final embeddings are used to discover SHS nodes in the network. GNN-SHS aims to discover SHSs in dynamic networks by reducing the computational time while achieving high accuracy. \textit{Our experimental results show that the proposed GNN-SHS model is at least 31.8 times faster and up to 2996.9 times faster than the baseline, providing a considerable efficiency advantage in run time.}

\noindent Our contributions are summarized as follows:
\begin{itemize}
\item \textbf{GNN-SHS model.} We propose an efficient graph neural network-based model GNN-SHS, for discovering SHSs in dynamic networks. %We consider the dynamic network as a sequence of snapshots of graph. 
Our model consider both the incremental and decremental edge updates in the network. GNN-SHS model preserves the network structure and node feature, and uses the final node embedding to discover SHSs.

\item \textbf{Theoretical analysis.} We theoretically show that the depth of the proposed graph neural network-based model GNN-SHS should be at least $\Omega({n}^2/\log^2 n)$ to solve the SHSs problem.

\item \textbf{Experimental analysis.} We validate the performance of the proposed GNN-SHS model on various synthetic and real-world datasets. The results demonstrate that on synthetic datasets, GNN-SHS is at least 31.8 times faster and, on average, 671.6 times faster than the baseline algorithm.
\end{itemize}

\textbf{Organization.} Section II discusses the related work. Section III presents the preliminaries and problem definition. Section IV discusses the proposed model GNN-SHS. Section V reports and discusses the extensive empirical results. Section VI concludes the paper.

\section{RELATED WORK} 
SHSs have numerous applications, such as community detection \cite{zhang2016social}, opinion control \cite{kuhlman2013controlling}, information diffusion \cite{sasabe2020mathematical}, viral marketing \cite{castiglione2020cognitive}, etc. Various studies \cite{lou2013mining, he2016joint, xu2019identifying, tang2012inferring, rezvani2015identifying, xu2017efficient, ding2016method, zhang2020finding, luo2020detecting, goel2021maintenance, goel2023enhancing} have been conducted to discover SHS nodes. Existing work on SHSs discovery can be categorized into 1) discovering SHSs in static networks; 2) discovering SHSs in dynamic networks. This section discusses the state-of-the-art for discovering SHSs. %Table \ref{related work} presents the summary of solutions for discovering structural hole spanners.

\subsection{Discovering SHSs in static networks} 
Lou et al. \cite{lou2013mining} proposed a technique based on minimal cut to discover SHSs in the network. However, the technique requires community information in the network, and the quality of identified SHS depends on how accurately are communities identified. He et al. \cite{he2016joint} developed a solution to identify SHSs and communities in the network. The authors explored the types of interactions, particularly for bridging nodes, in order to distinguish SHSs from the other nodes. Besides, the model assumed that every node is associated with one community only; however, in the real world, a node may belong to multiple communities \cite{yang2015defining}. Xu et al. \cite{xu2019identifying} proposed an algorithm to discover SHSs and argued that SHSs are the nodes that block maximum information propagation when removed from the network. Rezvani et al. \cite{rezvani2015identifying} developed a solution based on closeness centrality to discover SHSs. Xu et al. \cite{xu2017efficient} proposed various scalable techniques for discovering SHSs. Ding et al. \cite{ding2016method} devised a V-Constraint technique for identifying critical nodes that fill SH in the network. The model considered various node features such as the degree of the neighbour and topological features. However, the local features of a node may not capture the global importance of the SHS node. Tang et al. \cite{tang2012inferring} developed a solution based on shortest paths of length two to discover SHSs. Zhang et al. \cite{zhang2020finding} used a community forest model to discover SHS nodes in the network. The authors argued that local features-based metrics could not identify SHS nodes in the network. Luo et al. \cite{luo2020detecting} designed a deep learning model ComSHAE that identifies SHSs and communities in the network. The neural network based-model learns eigenvectors to infer SHS nodes and communities.

Numerous solutions have been proposed to discover SHSs for the steady-state behaviour of the network. However, real-world networks are not static; they evolve continuously.

\subsection{Discovering SHSs in dynamic networks} Goel et al. \cite{goel2021maintenance} designed a decremental algorithm to efficiently update SHSs in dynamic networks. The authors have considered one decremental edge update at a time. The algorithm first identifies a set of affected nodes due to changes in the network and then performs computations only on affected nodes. In contrast, we design a GNN-SHS model to discover SHSs in dynamic networks. Besides, we have considered both decremental and incremental edge updates in the network.

\section{PRELIMINARIES AND PROBLEM DEFINITION}
This section first discusses the preliminaries and background of the problem. Later, we formally state our SHS problem for the dynamic network.

\subsection{Preliminaries and background}
\noindent \textbf{Notations.} 
A graph can be defined as $G = (V, E)$, where $V$ is the set of nodes (vertices), and $E \subseteq V \times V$ is the set of edges. Let $n = |V |$ and $m = |E|$. We have considered unweighted and undirected graphs. A dynamic graph $G$ is modelled as a finite sequence of graphs $(G_t, G_{t+1}, ..., G_T)$. Each $G_t$ graph represents the network's state at a discrete-time interval $t$. We refer to each of the graph in the sequence as a snapshot graph. Each snapshot consists of the same set of nodes, whereas edges may appear or disappear over time. Hence, each graph snapshot can be described as an undirected graph $G_t = (V, E_t)$, containing all nodes and only alive edges at the time interval under consideration. Due to the dynamic nature of the graph, the edges in the graph may appear or disappear, due to which the label of the nodes (SHS or normal node) may change. Therefore, we need to design a technique that can discover SHSs in each new snapshot graph quickly.

Let $\vec{x}(i)$ denotes the feature vector of node $i$, $N(i)$ denotes the neighbors of node $i$, $h^{l}(i)$ represents the embedding of node $i$ at the $l^{th}$ layer of the model. $l$ represents the index of the aggregation layer, where $l = (1,2,...,L)$. Table \ref{notations} presents the symbols used in this paper.

\begin{definition}
\noindent \textbf{Pairwise connectivity \cite{goel2021maintenance}.} \normalfont \textit{The pairwise connectivity $u(i,j)$ for any node pair $(i,j)\in V \times V $ is computed as: }
{\begin{equation}
u(i,j)=\begin{cases} 1 & \text{if\,$i$\,and\,$j$\,are\,connected} \\ 0 & \text{otherwise}
\end{cases} 
\end{equation}}
\end{definition}

\begin{definition}
\noindent  \textbf{Total pairwise connectivity  \cite{goel2021maintenance}.}  \normalfont \textit{The total pairwise connectivity $P(G)$ is the pairwise connectivity across all pair of nodes in the graph and is computed as:}
{\begin{equation}
\label{eq: tpc}
\text{\ensuremath{{\displaystyle P(G)=\sum_{i,j\in V\times V,i\neq j}{\textstyle u(i,j)}}}}
\end{equation}}
\end{definition}

\noindent The\textit{ pairwise connectivity score \cite{goel2021maintenance}} $c(j)$ of node $j$ is its input to the total pairwise connectivity of the graph and is computed as:
\begin{equation}
\label{eq:score}
c(j)=P(G)-P(G\backslash\{j\})
\end{equation}
where $P(G)$ can be computed using Equation (\ref{eq: tpc}).

\begin{table}[t!] 
\caption{Table of symbols}
\label{notations}
\renewcommand{\arraystretch}{1.2}
\centering 
\begin{tabular}{ll} \hlineB{1.5} 
\textbf{Symbol} & \textbf{Definition} \\ \hlineB{1.5}
$G$ & Original graph \\ 
$V,E$ & Set of nodes and edges \\ 
$n,m$ & Number of nodes and edges\\
$k$ & Number of SHSs \\
$G_t$ & Snapshot of graph at time $t$ \\
$P(G)$ & Pairwise connectivity of graph \\ 
$P(G\backslash\{i\})$ & Pairwise connectivity of graph without node $i$ \\
$N(i)$ & Neighbors of node $i$ \\
$h^{l}{(i)}$ & Embedding of node $i$ at the $l^{th}$ layer \\ 
$||$ & Concatenation operator \\
$\sigma$ & Non-linearity \\ 
$l$ & Index of aggregation layer \\ 
$L$ & Total aggregation layers \\
$z(i)$ & Final embedding of node $i$ \\
$y(i)$ & Label of node $i$ \\
$\vec{x}(i)$ & Feature vector of node $i$\\ 
\hlineB{1.5}
\end{tabular} 
\end{table}

\vspace{0.05in}
\noindent \textbf{Graph Neural Networks (GNNs).} GNNs \cite{wu2020comprehensive} are designed by extending the deep learning methods for the graph data and are broadly utilized in various fields, e.g., computer vision, graph mining problems, etc. GNNs usually consist of graph convolution layers that extract local structural features of the nodes  \cite{velivckovic2017graph}. GNNs learn the representation of nodes by aggregating features from the ego network of node. Every node in network is described by its own features, and features of its neighbors \cite{kipf2016semi}.

\begin{figure}[ht!]
 \centering
 \includegraphics[width=0.5\columnwidth]{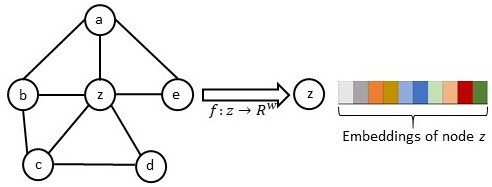}
 \caption{Embedding of node $i$.}
 \label{fig:embed}
\end{figure}

\vspace{0.05in}
\noindent \textbf{Network Embedding.} Network embedding \cite{cui2018survey} is a procedure with which network nodes can be described in a low-dimensional vector. Embedding intends to encode the nodes so that the resemblance in the embedding, approximates the resemblance in the network \cite{aguilar2021novel}. Embedding can be used for various graph mining problems, including node classification, regression problems, graph classification, etc. Figure \ref{fig:embed} depicts an example of node embedding in low-dimensional space.

\subsection{Problem Definition}
\noindent \textbf{Structural hole spanner problem for static network.} 

\noindent \textbf{\textit{Given:}} Given a graph $G = (V, E)$ and integer $k > 0$.

\noindent \textbf{\textit{Goal:}} \textit{Structural hole spanner problem} aims to discover a set of $k$ nodes called SHSs in $G$, where SHSs $\subset V$ and $|SHSs| = k$, such that when these nodes are removed from G, the total pairwise connectivity of the subgraph $ G(V\backslash SHSs)$ is minimized.

\begin{equation}
\text{SHSs} = {min}\,\,\{{P(G\backslash \text{SHSs})}\}
\end{equation}
where SHSs$\,\subset\,V$ and $|SHSs|=k$.

\noindent \textit{We refer to the $k$ nodes with the highest pairwise connectivity score in the graph as \textbf{top-\textbf{\textit{k}} SHSs} and rest as normal nodes.}

\noindent \textbf{Structural hole spanner problem for dynamic network.}

\noindent We represent the dynamic network as a sequence of snapshots of graph $(G_t, G_{t+1}, ..., G_T)$ and each snapshot graph describes all the edges that occur between a specified discrete-time interval (e.g., sec, minute, hour). Figure \ref{fig:evolving} illustrates four snapshots of graph taken at time $t, t+1, t+2$, $t+3$. Due to the dynamic nature of the graph, SHSs in the graph also change, and it is crucial to discover SHSs in each new snapshot graph quickly. Traditional SHSs techniques are time-consuming and may not be suitable for dynamic graphs. Therefore, we propose a GNN-based model to discover SHSs in dynamic networks by transforming the SHS discovery problem into a learning problem. We formally define the structural hole spanner problem for dynamic networks as follow:

\begin{figure}[ht!]
 \centering
 \includegraphics[width=0.5\columnwidth]{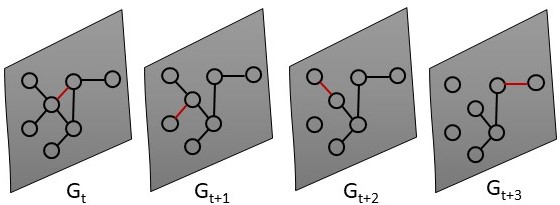}
 \caption{Illustration of snapshots of graph.}
 \label{fig:evolving}
\end{figure}

\noindent \textbf{{Given:}} Given snapshots of graph $G_t = (V,E_t), G_{t+1} = (V,E_{t+1})...., G_{T} = (V,E_{T})$ and integer $k > 0$.

\noindent \textbf{\textit{Goal:}} Train a model GNN-SHS to discover a set of $k$ SHSs in dynamic network (snapshots of graph). We aim to utilize the pre-trained model to discover SHSs in each new snapshot of the graph quickly. 

\vspace{0.1in}
\begin{theorem} [\textbf{Dinh et al. \cite{dinh2011new}}] \label{thm1}
\textit{\textbf{Discovering SHS problem is NP-hard.}}
\end{theorem}
\vspace{-0.1in}
\begin{proof}
We present an alternative proof, where we reduce the SHS model to vertex cover instead of $\beta$-Vertex Disruptor used in Dinh et al. \cite{dinh2011new}. The reason for this alternative proof is that it will be used as a foundation for Theorem 2. 

We show that the Vertex Cover (VC) problem is reducible to the SHS discovery problem. The definition of \textit{Structural Hole Spanners} states that SHSs are the set of $k$ nodes,  which, when deleted from the graph, minimizes the total pairwise connectivity of the subgraph. Let $G=(V, E)$ be an instance of a VC problem in an undirected graph $G$ with $V$ vertices and $E$ edges. \textit{VC problem} aims to discover a set of vertices of size $k$ such that the set includes at least one endpoint of every edge of the graph. If we delete the nodes in vertex cover from the graph, there will be no edge in the graph, and the pairwise connectivity of the residual graph will become $0$, i.e., $P(G)$ is minimized. In this way, we can say that graph $G$ has a VC of size $k$ if and only if graph $G$ have structural hole spanners of size $k$. Therefore, discovering the exact $k$ SHSs problem is NP-hard as a similar instance of a vertex cover problem is a known NP-hard problem.
\end{proof}
\vspace{0.1in}
\begin{theorem} \label{thm2}
\textit{\textbf{The depth of the proposed graph neural network-based GNN-SHS model with \textit{width} $=O(1)$ should be at least $\Omega({n}^2/\log^2 n)$ to solve the SHSs problem.} 
%(discover set of nodes removal of which minimises the total pairwise connectivity of the residual graph).
}
\end{theorem}
\vspace{-0.1in}
\begin{proof}
In Theorem \ref{thm1}, we proved that if we can discover SHS nodes in the graph, then we can solve the VC problem. Corollary 4.4 of Loukas \cite{loukas2019graph} showed that for solving the minimum VC problem, a message-passing GNN with a \textit{ width} $=O(1)$  should have a depth of at least $\Omega({n}^2/\log^2 n)$. Therefore, the lower bound on the depth of the VC problem also applies to our SHSs problem. Here, the \textit{depth} describes the number of layers in the GNN-SHS model, and \textit{width} indicates the number of hidden units.
\end{proof}
\vspace{0.1in}

Theorem 1 showed that discovering the exact $k$ SHSs in the network is an NP-hard problem. %Therefore, a shallow GNN can not be used to learn the exact SHSs due to the limitations on the depth of the model \cite{loukas2019graph}. 
Theorem 2 showed that the depth of GNN-SHS model should be at least $\Omega({n}^2/\log^2 n)$ to solve the SHSs problem. However, in practice, deeper GNN faces over-smoothing issue \cite{li2018deeper}, which results in poor model performance.  Due to the aforementioned concerns, we follow the same model adopted in Goel et al. \cite{goel2021maintenance}, where the authors settled for a greedy heuristic for finding the top-$k$ SHS nodes. Under the greedy algorithm, a node $j$ with maximum pairwise connectivity score (Equation \ref{eq:score}) is selected; this is the node which when removed from the network minimizes the total pairwise connectivity of the residual network. The node $j$ is then removed from the network and added to the SHS set. The process repeats until $k$ nodes are identified. It should be noted that this greedy algorithm, despite being a heuristic, is still computationally expensive with a complexity of  $O(kn(m+n))$. Nevertheless, real-world networks such as online social networks are dynamic, and they change rapidly. Since these networks change quickly,  the top-$k$ SHSs in the network also change continuously. A run time of $O(kn(m+n))$ of the greedy algorithm is too high and not suitable for practical purposes where speed is the key. For instance, it is highly possible that the network might already change by the time greedy algorithm computes the top-$k$ SHSs. Therefore, we need an efficient solution that can quickly discover top-$k$ SHSs in the changing networks.

The main idea of our approach is to rely on the greedy heuristic and treat its results as true labels to train a graph neural network for identifying the top-$k$ SHSs. The end result is a significantly faster heuristic for identifying the top-$k$ SHSs. Our heuristic is faster because we only have to train our graph neural network model once, and thereafter, whenever the graph changes the trained model can be utilized to discover SHSs.

\section{PROPOSED MODEL}\label{sec5}
%In this paper, we consider the dynamic network as a series of snapshots of graph, $G_t , G_{t+1}. . . , G_{T}$. Each snapshot describes the status of the network at time $t$. In addition, we consider that the nodes of the graph remain the same, whereas the edges may be added or removed in snapshots across the time intervals. Each graph snapshot can be described as an undirected graph $G_t = (V, E_t)$, containing all nodes and only alive edges at a given time interval under consideration. Figure \hl{} illustrates a network containing four snapshots of graph taken at time $t$ to $t+3$.

%\noindent Identifying SHS in dynamic networks is a challenging task. Nevertheless, the algorithm proposed by Goel et al. \cite{goel2021maintenance} might not work efficiently for real-world graphs that change drastically. Besides, the authors only consider one decremental update in the network. 

%In this section, we discuss the proposed model that discovers SHSs in a dynamic network. 
Inspired by the recent advancement of graph neural network techniques on various graph mining problems, we propose \textbf{\textit{GNN-SHS}}, a graph neural network-based framework to discover top-$k$ SHS nodes in the dynamic network. Figure \ref{fig:gnn_arch} represents the architecture of the proposed GNN-SHS model. We divided the SHSs identification process into two parts, i.e., model training and model application. The details of the GNN-SHS model are discussed below.

\subsection{Model Training} \label{model_train}
This section discusses the architecture of the proposed model and the training procedure.

\subsubsection{Architecture of GNN-SHS} 
In order to discover SHS nodes in dynamic network, we first transform the SHSs identification problem into a learning problem. We then propose a GNN-SHS model that uses the network structure as well as node features to identify SHS nodes. Our model utilizes three-node features, i.e., effective size \cite{burt1992structural}, efficiency \cite{burt1992structural} and degree, to characterize each node. These features are extracted from the one-hop ego network of the node. 

Given a graph and node features as input, our proposed model GNN-SHS first computes the low-dimensional node embedding vector and then uses the embedding of the nodes to determine the label of nodes (as shown in Figure \ref{fig:gnn_arch}). The label of a node can either be SHS or normal. The procedure for generating embeddings of the nodes is presented in Algorithm \ref{gnn-algo}. The model training is further divided into two phases: 1) Neighborhood Aggregation, 2) High Order Propagation. The two phases of the GNN-SHS model are discussed below:
\begin{figure*}[t!]
 \centering
 \includegraphics[width=0.63\paperwidth]{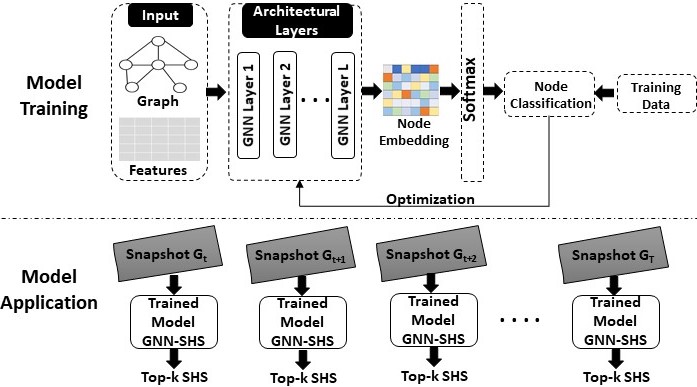}
 \caption{Architecture of proposed model GNN-SHS.}
 \label{fig:gnn_arch}
\end{figure*}
\noindent \textbf{Neighborhood Aggregation.} 
The neighborhood aggregation phase aggregates the features from the neighbors of a node to generate node embeddings. The node embeddings are the low dimensional representation of a node. Due to distinguishing characteristics exhibited by the SHSs, we considered all the one-hop neighbors of the node to create embedding. We generate the embeddings of node $i$ by aggregating the embeddings from its neighboring nodes, and we use the number of neighbors of node $i$ as weight factor:
\begin{equation}
\label{eq:agg}
h^{l}{(N(i))} = \sum_{j\in N(i)}{\frac{h^{l-1}{(j)}}{ \mid N(i) \mid }}
\end{equation}
where $h^{l}{(N(i))}$ represents the embedding vectors captured from the neighbors of node $i$. Embedding vector of each node is updated after aggregating embeddings from its neighbors. Node embeddings at layer $0$, i.e., $h^{0}(i)$ are initialized with the feature vectors $\vec{x}(i)$ of the nodes, i.e., Effective size, Efficiency and Degree. Each node retains its own feature information by concatenating its embedding vector from the previous layer with the aggregated embedding of its neighbors from the current layer as:
\begin{equation}
\label{eq:comb}
h^{l}{(i)} = \sigma{\Big(W^{l}\big(h^{l-1}{(i)} \mathbin\Vert h^{l}{(N(i))}\big) \Big)}
\end{equation}
where $W^{l}$ are the training parameter, $\mathbin\Vert$ is the concatenation operator, and $\sigma$ is the non-linearity, e.g., ReLU.

\noindent \textbf{High Order Propagation.} Our model employs multiple neighborhood aggregation layers in order to capture features from $l$-hop neighbors of a node. The output from the previous layer acts as input for the current layer. Stacking multiple layers will recursively form the representation $h^{l}(i)$ for node $i$ at the end of layer $l^{th}$ as: 
\begin{equation*}
z{(i)} = h^{l}{(i)}, \,\,\,\, \forall i \in V
\end{equation*}
where $z(i)$ denotes the final node embedding at the end of $l^{th}$ layer. For the purpose of classifying the nodes as SHS or normal node, we pass the final embeddings of each node $z{(i)}$ through the Softmax layer. This layer takes node embeddings as input and generates the probability of two classes: SHS and normal. We then train the model to distinguish between SHS and normal nodes.

\begin{algorithm}[ht!]
\caption{Generating node embeddings for GNN-SHS}
 \label{gnn-algo}
 \begin{algorithmic}[1]
 \renewcommand{\algorithmicrequire}{\textbf{Input:}}
 \renewcommand{\algorithmicensure}{\textbf{Output:}}
 \REQUIRE Graph $G(V,E)$, Input features $\vec{x}(i),\,\, \forall i \in V$, Depth $L$, Weight matrices $W^{l},\,\, \forall l \in \{1,..,L\}$, Non-linearity $\sigma$
\ENSURE Node embeddings $z{(i)}, \,\, \forall i \in V$ 
 %\\ \textit{Initialisation}:
 \STATE $h^{0}(i) \leftarrow \vec{x}(i),\,\,\forall i \in V$ 
 \FOR{$l = 1$ to $L$}
 \FOR{$i \in V$}
  \STATE Compute $h^{l}{(N(i))}$ using Equation (\ref{eq:agg})
  \STATE Compute $h^{l}{(i)}$ using Equation (\ref{eq:comb})
 \ENDFOR
 \ENDFOR
 \STATE $z{(i)} = h^{L}{(i)}$
\end{algorithmic}
\end{algorithm}

\subsubsection{Training procedure} To discover SHS nodes in the network, we employ \textit{binary cross-entropy loss} to train the model with the actual label known for a set of nodes. The loss function $\mathcal{L}$ is computed as:

\begin{equation*}
\label{eq_loss}
\mathcal{L} = {-}\frac{1}{ r}{\sum_{i=1}^{r}\bigg(y(i)\log{\hat{y}(i)} + (1-y(i)) \log{(1-\hat{y}(i))\bigg)}}
\end{equation*}
where $y$ is the true label of a node, $\hat{y}$ is the label predicted by the GNN-SHS model, and $r$ is the number of nodes in the training data for which the labels are known.

\subsection{Model Application}
We first train the GNN-SHS model using the labelled data as discussed in Section \ref{model_train}. Once the model is trained, we then utilize the trained GNN-SHS model to quickly discover SHS nodes in each snapshot of graph (snapshot obtained from the dynamic network). The discovered SHSs are nodes removal of which minimizes the total pairwise connectivity in the remaining subgraph.

\section{EXPERIMENTS} \label{exp}
We implemented the code in Python on a machine with i7-8700 CPU. %This section reports the performance of the proposed model GNN-SHS by performing extensive experiments on various datasets. 
This section first discusses the datasets used for experimental analysis, followed by evaluation metrics. We then discuss the baseline, ground truth computation, and training settings. Lastly, we report and discuss the results obtained.

\subsection{Datasets}
The details of the synthetic and real-world datasets are discussed below.

\noindent \textbf{\textit{Synthetic Datasets}}. We analyze the performance of GNN-SHS model by conducting experiments on synthetic datasets. We generate synthetic networks using graph-generating algorithms and vary the network size to determine its effect on model performance. We conduct experiments on synthetic networks with diverse topologies: Preferential Attachment (PA) networks and Erdos-Renyi (ER) \cite{erdHos1959random} networks. We generate PA$(n)$ network with 500, 1000 and 1500 nodes, where $n$ denotes total nodes in the network. For ER$(n,p)$, we generate networks with 250 and 500 nodes, where $p$ is the probability of adding an edge to the network. In PA$(n)$ network, a highly connected node is more likely to get new neighbors. In ER$(n,p)$ network, parameter $p$ acts as a weighting function, and there are higher chances that the graph contains more edges as $p$ increases from 0 to 1. The properties of synthetic datasets are presented in Table \ref{dataset2}.

\noindent \textbf{\textit{Real-World Datasets}}. 
We also conduct experiments on real-world datasets to analyze GNN-SHS performance. Properties of real-world datasets are presented in Table \ref{datasetreal}.

\begin{table}[ht!] 
\caption{Properties of synthetic datasets.}
\label{dataset2}
\renewcommand{\arraystretch}{1.4}
\centering 
\begin{tabular}{llll} \hlineB{1.5} 
\textbf{Dataset} & \textbf{Nodes} & \textbf{Edges} & \textbf{Average degree} \\ \hlineB{1.5}
PA (500) & 500 & 499 & 2 \\ 
PA (1000) & 1000 & 999 & 2\\ 
PA (1500) & 1500 & 1499 & 2\\ 
ER (250, 0.01) & 250 & 304 & 2\\ 
ER (250, 0.5) & 250 & 15583 & 124\\ 
ER (500, 0.04) & 500 & 512 & 2\\ 
ER (500, 0.5) & 500 & 62346 & 249\\ \hlineB{1.5} 
\end{tabular} 
\end{table}

\begin{table}[ht!] 
\caption{Properties of real-world datasets.}
\label{datasetreal}
\renewcommand{\arraystretch}{1.4}
\centering 
\begin{tabular}{llll} \hlineB{1.5} 
\textbf{Dataset} & \textbf{Type} & \textbf{Nodes} & \textbf{Edges} \\ \hlineB{1.5}
Dolphin \cite{lusseau2003bottlenose} & Social network & 62 & 159 \\ 
American College Football \cite{girvan2002community} & Football network & 115 & 613\\ 
 \hlineB{1.5} 
\end{tabular} 
\end{table}

\subsection{Evaluation Metrics}
We measure the \textit{efficiency} of our proposed GNN-SHS model in terms of speedup achieved by GNN-SHS over comparative methods. The speedup is computed as follows:
\begin{equation*}
\label{agg}
\text{Speedup} = \frac{\text{Run time of comparative method}}{\text{Run time of proposed GNN-SHS model}}
\end{equation*}

In addition, we measure the \textit{effectiveness} of our model in terms of classification accuracy achieved by the model.

\subsection{Baseline}
In the literature, there is only one decremental algorithm proposed by Goel et al. \cite{goel2021maintenance} that addresses the problem of identifying SHSs in dynamic networks. Therefore, we compare the performance of our model GNN-SHS with the decremental algorithm. The authors \cite{goel2021maintenance} in their work considered only a single edge deletion update at a time.

% However, in our model, we have considered deletion as well as additional edge updates.

\subsection{Ground Truth Computation} To compute the ground truth labels, we first calculate the connectivity score $c$ for each node and then label top-$k$ nodes with the highest score as SHS nodes and rest as normal nodes. For experimental analysis, we set the value of $k$ to 50. %Labelled nodes are used to train the model.

\subsection{Training Settings} We implemented the code of GNN-SHS in PyTorch and fixed the number of layers to 2 and the embedding dimension to 32. The model parameters are trained using Adam optimizer, learning rate of 0.01 and weight decay of $5e-4$. We train GNN-SHS model for 200 epochs. We used 60\% of the nodes for training, 20\% for validation, and 20\% for testing. In addition, we used an inductive setting where test nodes are unseen to the model during the training phase. 

\subsection{Performance of GNN-SHS}
In dynamic network, as the graph changes with time, we get multiple snapshots of graph. The trained model can be used to identify the SHSs in each snapshot of the dynamic graph. Even if it takes some time to train the model, we need to train it only once, and after that, whenever the graph changes, the trained GNN-SHS model can be used to discover the updated SHSs in a few seconds.

\subsubsection{Results on synthetic dataset}
The decremental algorithm \cite{goel2021maintenance} only works for a single edge deletion update. Therefore, we analyze our model GNN-SHS performance on a single edge deletion update only so that we can compare the speedup of GNN-SHS over decremental algorithm. To determine the speedup of our proposed GNN-SHS model over decremental algorithm, we start from the whole network and arbitrarily delete 50 edges; we only delete one edge at a time. In this way, we obtain multiple snapshots of graph. We set the value of $k$ (number of SHS) to 50 and make use of the trained GNN-SHS model to discover SHS nodes in each new snapshot graph. We calculate the geometric mean of the speedup achieved by GNN-SHS over the decremental algorithm. In our results, the geometric mean denotes the geometric mean of the speedup attained by our model, Min denotes the minimum speedup attained, and Max indicates the maximum speedup attained over the decremental algorithm.

\begin{table}[ht!] 
\caption{Classification accuracy of GNN-SHS on synthetic datasets.}
\label{ACC_GNN-SHS}
\renewcommand{\arraystretch}{1.3}
\centering 
\begin{tabular}{ll} \hlineB{1.5}
\textbf{Dataset} & \textbf{Accuracy}\\ \hlineB{1.5}
PA (500) & 94.5\%\\ 
PA (1000) & 95.33\% \\ 
PA (1500) & 96.5\%\\ 
ER (250, 0.01) & 86\%\\
ER (250, 0.5) & 90\%\\ 
ER (500, 0.04) & 87\%\\
ER (500, 0.5) & 92\%\\ \hlineB{1.5}
\end{tabular} 
\end{table}

\begin{table}[ht!] 
\caption{Speedup of GNN-SHS on decremental algorithm \cite{goel2021maintenance} over 50 edge deletions on synthetic datasets.}
\label{SPEED_GNN-SHS}
\renewcommand{\arraystretch}{1.3}
\centering 
\begin{tabular}{llll} \hlineB{1.5}
\textbf{Dataset} & \textbf{Geometric Mean} & \textbf{Min} & \textbf{Max} \\ \hlineB{1.5}
PA (500) & 1236.4 & 1012.5 & 1532.7\\ 
PA (1000) & 1930.6 & 1574.2 & 2141.4\\ 
PA (1500) & 2639.7 & 2432.1 & 2996.9\\ 
ER (250, 0.01) & 37.5 & 31.8 & 40.2\\
ER (250, 0.5) & 287.3 & 263.6 & 301.2\\ 
ER (500, 0.04) & 368.2 & 354.5 & 379.3\\
ER (500, 0.5) & 2466.6 & 2015.3 & 2845.9\\ \hlineB{1.5}
\textbf{Mean (Geometric)} & 671.6 & 584.1 & 745.9\\ \hlineB{1.5}
\end{tabular} 
\end{table}

Table \ref{ACC_GNN-SHS} reports the classification accuracy (SHS detection accuracy) achieved by GNN-SHS on various synthetic graphs. GNN-SHS achieves a minimum accuracy of 94.5\% on Preferential Attachment graph PA(500) and 86\% classification accuracy for Erdos-Renyi graph ER(250, 0.01). The results from the table show that graph neural network-based models achieve high SHS classification accuracy. 

Table \ref{SPEED_GNN-SHS} reports the speedup achieved by the proposed GNN-SHS model over the decremental algorithm. %The minimum speedup attained by the GNN-SHS model is 31.8 times for ER(250, 0.01) network, and the maximum speedup attained is 2996.9 times for PA(1500) network. 
\textit{Our model achieved high speedup over the decremental algorithm while sacrificing a small amount of accuracy.} \textit{The proposed model is at least 31.8 times faster for ER(250, 0.01) network and up to 2996.9 times faster for PA(1500) over the decremental algorithm, providing a considerable efficiency advantage.} The geometric mean speedup is always at least 37.5 times, and the average speedup over all tested datasets is 671.6 times. Results show that our graph neural network-based model GNN-SHS speeds up the SHS identification process in dynamic networks. In addition, it has been observed from the results that the speedup increases as network size increases, e.g., for PA graphs, the geometric mean speedup is 1236.4 times for a graph of 500 nodes, 1930.6 times for a graph with 1000 nodes and 2639.7 times for a graph with 1500 nodes.

In Theorem 2, we showed that the depth of GNN-SHS should be at least $\Omega({n}^2/\log^2 n)$ to solve the SHSs problem. Nevertheless, a deeper graph neural network suffers from an over-smoothing problem \cite{li2018deeper, yang2020toward}, making it challenging for GNN-SHS to differentiate between the embeddings of the nodes. In order to avoid the over-smoothing problem, we only used 2 layers in our GNN-SHS model. %The model performance starts deteriorating by increasing the number of layers in the model.\\

\subsubsection{Results on real-world dataset}
We perform experiments on real-world datasets to determine the proposed model performance for incremental and decremental batch updates. In the literature, no solution discovers SHSs for incremental and decremental batch updates; therefore, we can not compare our results with other solutions. We only report the results obtained from our experiments. We set the value of $k=5$ (number of SHSs). For each real-world dataset, we initiate with the whole network and then arbitrarily delete 5 edges from the network and add 5 edges to the network at once. In this manner, we obtain a snapshot of the graph. We then use our trained model to discover SHSs in the new snapshot graph. Our empirical results in Table \ref{result-real} show that our model discovers updated SHSs in less than 1 second for both Dolphin and American College Football datasets. Besides, our model achieves high classification accuracy in discovering SHSs for batch updates.

\begin{table}[ht!] 
\caption{Run time (sec) and classification accuracy of GNN-SHS on real-world datasets.}
\label{result-real}
\renewcommand{\arraystretch}{1.4}
\centering 
\begin{tabular}{p{3.6cm}ll} \hlineB{1.5} 
\textbf{Dataset} & \textbf{Run time (sec)}& \textbf{Accuracy} \\ \hlineB{1.5}
Dolphin \cite{lusseau2003bottlenose} & 0.002 & 76.92\% \\ 
American College Football \cite{girvan2002community} & 0.009 & 86.96\%\\ 
 \hlineB{1.5} 
\end{tabular} 
\end{table}

\begin{comment}
\begin{itemize}
  \item Dolphins \cite{lusseau2003bottlenose} is a social network, which consists of 62 nodes and 159 edges.
  \item American College Football \cite{girvan2002community} is a football network which consists of 115 nodes and 613 edges. \\
\end{itemize}
\end{comment}

\section{CONCLUSION}
The structural hole spanner identification problem has various applications, including community detection, viral marketing, etc. However, the problem has not been studied well for dynamic networks where nodes and edges can join or leave the network. This paper studied the problem of discovering SHSs in dynamic networks. We considered the dynamic network as a sequence of snapshots and proposed GNN-SHS model, that discovers SHSs in the dynamic network by learning low-dimensional embedding vectors of the nodes. We performed empirical analysis on various datasets. Our results show that the proposed GNN-SHS model is at least 31.8 times faster than the comparative method, demonstrating a considerable advantage in run time. 

\bibliographystyle{unsrtnat}
\bibliography{main}

\begin{thebibliography}{33}
\providecommand{\natexlab}[1]{#1}
\providecommand{\url}[1]{\texttt{#1}}
\expandafter\ifx\csname urlstyle\endcsname\relax
  \providecommand{\doi}[1]{doi: #1}\else
  \providecommand{\doi}{doi: \begingroup \urlstyle{rm}\Url}\fi

\bibitem[Zannettou et~al.(2018)Zannettou, Caulfield, Blackburn, De~Cristofaro,
  Sirivianos, Stringhini, and Suarez-Tangil]{zannettou2018origins}
Savvas Zannettou, Tristan Caulfield, Jeremy Blackburn, Emiliano De~Cristofaro,
  Michael Sirivianos, Gianluca Stringhini, and Guillermo Suarez-Tangil.
\newblock On the origins of memes by means of fringe web communities.
\newblock In \emph{Proceedings of the Internet Measurement Conference 2018},
  pages 188--202, 2018.
\newblock \doi{https://doi.org/10.1145/3278532.3278550}.

\bibitem[Burt(2009)]{burt2009structural}
Ronald~S Burt.
\newblock \emph{Structural holes: The social structure of competition}.
\newblock Harvard university press, 2009.

\bibitem[Lou and Tang(2013)]{lou2013mining}
Tiancheng Lou and Jie Tang.
\newblock Mining structural hole spanners through information diffusion in
  social networks.
\newblock In \emph{WWW}, pages 825--836, 2013.

\bibitem[He et~al.(2016)He, Lu, Ma, Cao, Shen, and Yu]{he2016joint}
Lifang He, Chun-Ta Lu, Jiaqi Ma, Jianping Cao, Linlin Shen, and Philip~S Yu.
\newblock Joint community and structural hole spanner detection via harmonic
  modularity.
\newblock In \emph{Proceedings of the 22nd ACM SIGKDD International Conference
  on Knowledge Discovery and Data Mining}, pages 875--884, 2016.
\newblock \doi{https://doi.org/10.1145/2939672.2939807}.

\bibitem[Xu et~al.(2019)Xu, Li, Liang, Yu, Yang, and Gao]{xu2019identifying}
Wenzheng Xu, Tong Li, Weifa Liang, Jeffrey~Xu Yu, Ning Yang, and Shaobing Gao.
\newblock Identifying structural hole spanners to maximally block information
  propagation.
\newblock \emph{Information Sciences}, 505:\penalty0 100--126, 2019.

\bibitem[Tang et~al.(2012)Tang, Lou, and Kleinberg]{tang2012inferring}
Jie Tang, Tiancheng Lou, and Jon Kleinberg.
\newblock Inferring social ties across heterogenous networks.
\newblock In \emph{Proceedings of the fifth ACM international conference on Web
  search and data mining}, pages 743--752, 2012.

\bibitem[Rezvani et~al.(2015)Rezvani, Liang, Xu, and
  Liu]{rezvani2015identifying}
Mojtaba Rezvani, Weifa Liang, Wenzheng Xu, and Chengfei Liu.
\newblock Identifying top-k structural hole spanners in large-scale social
  networks.
\newblock In \emph{CIKM}, pages 263--272, 2015.

\bibitem[Xu et~al.(2017)Xu, Rezvani, Liang, Yu, and Liu]{xu2017efficient}
Wenzheng Xu, Mojtaba Rezvani, Weifa Liang, Jeffrey~Xu Yu, and Chengfei Liu.
\newblock Efficient algorithms for the identification of top-$ k $ structural
  hole spanners in large social networks.
\newblock \emph{TKDE}, 29\penalty0 (5):\penalty0 1017--1030, 2017.

\bibitem[Ding et~al.(2016)Ding, Wang, and Wei]{ding2016method}
Liting Ding, Jun Wang, and Wei Wei.
\newblock Method for detecting key nodes who occupy structural holes in social
  network sites.
\newblock In \emph{Pacific Asia Conference On Information Systems (PACIS)}.
  Association For Information System, 2016.

\bibitem[Zhang et~al.(2020)Zhang, Xu, Xu, Deng, Gu, Ma, Lai, Hu, Yu, Hou,
  et~al.]{zhang2020finding}
Yan Zhang, Hua Xu, Yunfeng Xu, Junhui Deng, Juan Gu, Rui Ma, Jie Lai, Jiangtao
  Hu, Xiaoshuai Yu, Lei Hou, et~al.
\newblock Finding structural hole spanners based on community forest model and
  diminishing marginal utility in large scale social networks.
\newblock \emph{Knowledge-Based Systems}, 199:\penalty0 105916, 2020.
\newblock \doi{https://doi.org/10.1016/j.knosys.2020.105916}.

\bibitem[Luo and Du(2020)]{luo2020detecting}
JiaXing Luo and YaJun Du.
\newblock Detecting community structure and structural hole spanner
  simultaneously by using graph convolutional network based auto-encoder.
\newblock \emph{Neurocomputing}, 410:\penalty0 138--150, 2020.
\newblock \doi{https://doi.org/10.1016/j.neucom.2020.05.039}.

\bibitem[Goel et~al.(2021)Goel, Shen, Tian, and Guo]{goel2021maintenance}
Diksha Goel, Hong Shen, Hui Tian, and Mingyu Guo.
\newblock Maintenance of structural hole spanners in dynamic networks.
\newblock In \emph{2021 IEEE 46th Conference on Local Computer Networks (LCN)},
  pages 339--342. IEEE, 2021.
\newblock \doi{https://doi.org/10.1109/LCN52139.2021.9524948}.

\bibitem[Goel et~al.(2024)Goel, Shen, Tian, and Guo]{goel2024effective}
Diksha Goel, Hong Shen, Hui Tian, and Mingyu Guo.
\newblock Effective graph-neural-network based models for discovering
  structural hole spanners in large-scale and diverse networks.
\newblock \emph{Expert Systems with Applications}, 249:\penalty0 123636, 2024.

\bibitem[Thekumparampil et~al.(2018)Thekumparampil, Wang, Oh, and
  Li]{thekumparampil2018attention}
Kiran~K Thekumparampil, Chong Wang, Sewoong Oh, and Li-Jia Li.
\newblock Attention-based graph neural network for semi-supervised learning.
\newblock \emph{arXiv preprint arXiv:1803.03735}, 2018.

\bibitem[Kipf and Welling(2017)]{kipf2016semi}
Thomas~N Kipf and Max Welling.
\newblock Semi-supervised classification with graph convolutional networks.
\newblock \emph{International Conference on Learning Representations (ICLR)},
  2017.

\bibitem[Zhang et~al.(2016)Zhang, Qiu, Guo, Guo, and Xiong]{zhang2016social}
Qishan Zhang, Qirong Qiu, Wenzhong Guo, Kun Guo, and Naixue Xiong.
\newblock A social community detection algorithm based on parallel grey label
  propagation.
\newblock \emph{Computer Networks}, 107:\penalty0 133--143, 2016.

\bibitem[Kuhlman et~al.(2013)Kuhlman, Kumar, and Ravi]{kuhlman2013controlling}
Chris~J Kuhlman, VS~Anil Kumar, and SS~Ravi.
\newblock Controlling opinion propagation in online networks.
\newblock \emph{Computer Networks}, 57\penalty0 (10):\penalty0 2121--2132,
  2013.

\bibitem[Sasabe(2020)]{sasabe2020mathematical}
Masahiro Sasabe.
\newblock Mathematical epidemiological analysis of dynamics of delay attacks on
  pull-based competitive information diffusion.
\newblock \emph{Computer Networks}, 180:\penalty0 107383, 2020.

\bibitem[Castiglione et~al.(2020)Castiglione, Cozzolino, Moscato, and
  Moscato]{castiglione2020cognitive}
Aniello Castiglione, Giovanni Cozzolino, Francesco Moscato, and Vincenzo
  Moscato.
\newblock Cognitive analysis in social networks for viral marketing.
\newblock \emph{IEEE Transactions on Industrial Informatics}, 2020.

\bibitem[Goel(2023)]{goel2023enhancing}
Diksha Goel.
\newblock Enhancing network resilience through machine learning-powered graph
  combinatorial optimization: Applications in cyber defense and information
  diffusion.
\newblock \emph{arXiv preprint arXiv:2310.10667}, 2023.

\bibitem[Yang and Leskovec(2015)]{yang2015defining}
Jaewon Yang and Jure Leskovec.
\newblock Defining and evaluating network communities based on ground-truth.
\newblock \emph{Knowledge and Information Systems}, 42\penalty0 (1):\penalty0
  181--213, 2015.
\newblock \doi{https://doi.org/10.1007/s10115-013-0693-z}.

\bibitem[Wu et~al.(2020)Wu, Pan, Chen, Long, Zhang, and
  Philip]{wu2020comprehensive}
Zonghan Wu, Shirui Pan, Fengwen Chen, Guodong Long, Chengqi Zhang, and S~Yu
  Philip.
\newblock A comprehensive survey on graph neural networks.
\newblock \emph{IEEE transactions on neural networks and learning systems},
  32\penalty0 (1):\penalty0 4--24, 2020.
\newblock \doi{https://doi.org/10.1109/TNNLS.2020.2978386}.

\bibitem[Veli{\v{c}}kovi{\'c} et~al.(2018)Veli{\v{c}}kovi{\'c}, Cucurull,
  Casanova, Romero, Lio, and Bengio]{velivckovic2017graph}
Petar Veli{\v{c}}kovi{\'c}, Guillem Cucurull, Arantxa Casanova, Adriana Romero,
  Pietro Lio, and Yoshua Bengio.
\newblock Graph attention networks.
\newblock \emph{International Conference on Learning Representations (ICLR)},
  2018.

\bibitem[Cui et~al.(2018)Cui, Wang, Pei, and Zhu]{cui2018survey}
Peng Cui, Xiao Wang, Jian Pei, and Wenwu Zhu.
\newblock A survey on network embedding.
\newblock \emph{IEEE transactions on knowledge and data engineering},
  31\penalty0 (5):\penalty0 833--852, 2018.

\bibitem[Aguilar-Fuster and Rubio-Loyola(2021)]{aguilar2021novel}
Christian Aguilar-Fuster and Javier Rubio-Loyola.
\newblock A novel evaluation function for higher acceptance rates and more
  profitable metaheuristic-based online virtual network embedding.
\newblock \emph{Computer Networks}, 195:\penalty0 108191, 2021.

\bibitem[Dinh et~al.(2011)Dinh, Xuan, Thai, Pardalos, and Znati]{dinh2011new}
Thang~N Dinh, Ying Xuan, My~T Thai, Panos~M Pardalos, and Taieb Znati.
\newblock On new approaches of assessing network vulnerability: hardness and
  approximation.
\newblock \emph{IEEE/ACM Transactions on Networking}, 20\penalty0 (2):\penalty0
  609--619, 2011.

\bibitem[Loukas(2020)]{loukas2019graph}
Andreas Loukas.
\newblock What graph neural networks cannot learn: depth vs width.
\newblock \emph{International Conference on Learning Representations (ICLR)},
  2020.

\bibitem[Li et~al.(2018)Li, Han, and Wu]{li2018deeper}
Qimai Li, Zhichao Han, and Xiao-Ming Wu.
\newblock Deeper insights into graph convolutional networks for semi-supervised
  learning.
\newblock In \emph{Thirty-Second AAAI conference on artificial intelligence},
  2018.

\bibitem[Burt(1992)]{burt1992structural}
R~Burt.
\newblock Structural holes: the social structure of competition.
\newblock \emph{Harvard, MA, Harvard University Press}, pages 23--100, 1992.

\bibitem[Erd\"os and R\'enyi(1959)]{erdHos1959random}
P~Erd\"os and A~R\'enyi.
\newblock On random graphs i.
\newblock \emph{Publicationes Mathematicae Debrecen}, 6:\penalty0 290--297,
  1959.

\bibitem[Lusseau et~al.(2003)Lusseau, Schneider, Boisseau, Haase, Slooten, and
  Dawson]{lusseau2003bottlenose}
David Lusseau, Karsten Schneider, Oliver~J Boisseau, Patti Haase, Elisabeth
  Slooten, and Steve~M Dawson.
\newblock The bottlenose dolphin community of doubtful sound features a large
  proportion of long-lasting associations.
\newblock \emph{Behavioral Ecology and Sociobiology}, 54\penalty0 (4):\penalty0
  396--405, 2003.
\newblock \doi{https://doi.org/10.1007/s00265-003-0651-y}.

\bibitem[Girvan and Newman(2002)]{girvan2002community}
Michelle Girvan and Mark~EJ Newman.
\newblock Community structure in social and biological networks.
\newblock \emph{Proceedings of the national academy of sciences}, 99\penalty0
  (12):\penalty0 7821--7826, 2002.
\newblock \doi{https://doi.org/10.1073/pnas.122653799}.

\bibitem[Yang et~al.(2020)Yang, Gu, Wang, Cao, Zhai, Jin, and
  Guo]{yang2020toward}
Liang Yang, Junhua Gu, Chuan Wang, Xiaochun Cao, Lu~Zhai, Di~Jin, and Yuanfang
  Guo.
\newblock Toward unsupervised graph neural network: Interactive clustering and
  embedding via optimal transport.
\newblock In \emph{2020 IEEE International Conference on Data Mining (ICDM)},
  pages 1358--1363. IEEE, 2020.

\end{thebibliography}

%\bibliographystyle{ACM-Reference-Format}
%\bibliography{sample-base}
\end{document}